\documentclass[reprint,amsmath,amssymb,aps,superscriptaddress,pre,floatfix]{revtex4-2}

\usepackage{graphicx}
\usepackage{xcolor}
\usepackage{dcolumn}
\usepackage{bm}
\usepackage{booktabs}
\usepackage{sidecap}
\usepackage{upgreek}
\usepackage{here}
\sidecaptionvpos{figure}{t}
\usepackage[pdftex,bookmarks,colorlinks,breaklinks]{hyperref} 
\hypersetup{linkcolor=red,citecolor=blue,filecolor=dullmagenta,urlcolor=blue} 
\hypersetup{pdftoolbar=true,pdfpagemode=UseNone,pdfstartview=FitH, colorlinks=true,extension=}

\begin{document}

\title[Double tunnel]{Enhanced intraband absorption in two-step photon upconversion solar cells with a double-tunnel-junction structure}
\author{Koichiro Matsuzawa}
\affiliation{Department of Electrical and Electronic Engineering, Graduate School of Engineering, Kobe University, 1-1 Rokkodai, Nada, Kobe 657-8501, Japan}

\author{Shigeo Asahi}
\email[]{asahis@people.kobe-u.ac.jp}
\affiliation{Department of Electrical and Electronic Engineering, Graduate School of Engineering, Kobe University, 1-1 Rokkodai, Nada, Kobe 657-8501, Japan}

\author{Takashi Kita}
\affiliation{Department of Electrical and Electronic Engineering, Graduate School of Engineering, Kobe University, 1-1 Rokkodai, Nada, Kobe 657-8501, Japan}

\date{\today}
\begin{abstract}
Two-step photon upconversion solar cells (TPU-SCs) belong to the class of solar cells that in principle can exceed the Shockley--Queisser limit for single-junction solar cells. A TPU-SC basically consists of a wide-gap and a narrow-gap semiconductor layer, where intraband absorption of infrared (IR) photons by electrons accumulated at the conduction-band-edge discontinuity leads to an increase in the output current and voltage. This IR-induced upconversion process enables a better utilization of the broad solar spectrum, and quantum dots (QDs) can be added to improve the intraband transition rates in actual devices. In this study, we added an n--p--n double-tunnel-junction structure to an $\mathrm{Al_{0.3}Ga_{0.7}As}$/GaAs TPU-SC including a QD layer to further enhance the contribution of intraband absorption to the output current. The double-tunnel-junction structure should suppress carrier recombination at the heterointerface and improve the extraction efficiency of electrons after intraband absorption. We performed two-color excitation experiments using IR and 784-nm light, and we particularly found that 784-nm light causes a significant amount of intraband transitions in this device. By employing rate equations, we clarify that this result demonstrates a higher intraband absorption efficiency due to the double-tunnel-junction structure. This highlights the potential of double-tunnel-junction structures for the realization of high-efficiency TPU-SCs.
\end{abstract}

\maketitle

\section{Introduction}\label{sec:Introduction}
The progress of solar cell (SC) technologies has led to significant improvements in terms of maximum energy conversion efficiency as well as cost performance \cite{Vos1980,Ross1982,Ekins-Daukes2001,Green2001,Okada2015,Lin2020,Green2025}. While single-junction SCs made of silicon constitute an important group in these technologies from the viewpoint of practical usage, their theoretical conversion efficiency is limited to 32\% under 1 sun illumination \cite{Hirst2011,Kita2019,Shockley1961}. This is mainly due to two types losses in this relatively simple design: thermal losses caused by photons with an energy larger than the bandgap energy of the absorber material, and transmission losses caused by photons with a smaller energy. In light of these fundamental limitations, several types of SCs with more complex device structures have been proposed and developed. The most successful type is the multi-junction SC comprising a series of single-junction SCs connected via tunnel junctions. Another type is the intermediate-band (IB) SC \cite{Luque1997}, where a so-called IB is added between the valence band (VB) and the conduction band (CB) of the absorber layer of a single-junction SC, and this can be achieved using quantum dots (QDs) or impurities \cite{mart2006,Okada2011,Hwang2014,Kada2015,Kada2017,Shoji2017,Sogabe2021}. In addition to the photocurrent generated by direct band-to-band transitions, transitions from the VB to the IB and then from the IB to the CB can occur in such a device, which allows us to use low-energy photons in the solar spectrum. These transitions from the VB to the CB via the IB can lead to an increase in the output current (accompanied by a slight voltage reduction); the maximum theoretical conversion efficiency of an IBSC under 1 sun illumination determined using the detailed balanced approach is 46.8\%. However, the actually achieved increases in current due to the transitions via the IB are usually too small to significantly increase the energy conversion efficiency. One reason for the small contribution of sub-bandgap photon absorption is the short lifetime of electrons in the IB (they recombine before they can reach the CB). To enhance the effect of sub-bandgap photon absorption, the photon ratchet IBSC concept has been proposed \cite{Sogabe2021,Yoshida2012}. In this type of SC, the electrons excited to the IB are immediately transported to a ratchet band radiatively decoupled from the VB, and thus the electrons have a longer lifetime. In other words, unlike in the case of the classic IBSC concept, the final state of the first transition step and the initial state of the second transition step (from the VB to the CB) are spatially separated.

Recently, we presented a particular type of photon-ratchet-based SC called two-step photon upconversion SC (TPU-SC) \cite{Asahi2017,Asahi2018,Watanabe2021,Kinugawa2019,Asahi2019}. As the name suggests, a TPU-SC basically relies on the idea of using two sequential transitions for improvement, without relying on an IB inside the bandgap region. It is a single-junction SC composed of a wide-bandgap semiconductor (WGS) and a narrow-bandgap semiconductor (NGS), where the WGS layer absorbs the high-energy photons and the NGS layer absorbs the remaining photons with energy larger than the NGS bandgap energy. The latter absorption process is the first step of the considered TPU process. Due to the internal electric field, the electrons in the NGS CB drift to the WGS/NGS heterointerface, where they accumulate to a certain degree. The second step of the TPU process is the photoexcitation of accumulated electrons to the WGS CB, which constitutes an intraband excitation process. The photon energy required for this second step is less than the NGS bandgap energy [\textit{e.g.}, infrared (IR) light can usually be used], and thus this process reduces the transmission losses and allows us to exceed the conversion efficiency limit for single-junction SCs. An interesting adaption of this design is the addition of QDs to relax the optical selection rule for intraband transition. Based on a theoretical calculation using the detailed balance approach, the maximum theoretical conversion efficiency of a TPU-SC with a band-edge offset ratio of 3:2 is 44.2\% under 1 sun illumination \cite{Asahi2018}. In a previous study, we demonstrated that the output current of an $\mathrm{Al_{0.3}Ga_{0.7}As}$/GaAs TPC-SC with InAs QDs can be increased by additional IR illumination \cite{Asahi2017,Watanabe2021}. The observed current increase was attributed to the photoexcitation of electrons from the NGS CB edge to the WGS CB, which enables more electrons to reach the top electrode. In addition to this current increase, we also confirmed an IR-induced enhancement in the open-circuit voltage (the upconverted electrons lead to an increase in the WGS electron quasi-Fermi level, which increases the output voltage). However, the achieved levels of the current increase and the accompanying voltage boost are not yet sufficient for applications. To further improve the performance, we recently proposed to add a double-tunnel-junction structure to the TPU-SC \cite{Matsuzawa2024}. We confirmed that a clear enhancement of the IR-induced voltage boost can be achieved. On the other hand, the details of the current generation process have not yet been investigated. 

In this work, we present experimental results that reveal enhanced photocurrent generation through intraband absorption in a TPU-SC with a double-tunnel-junction structure. In Section~\ref{sec:Experimental}, we explain the operating principle based on the calculated band alignment and provide the experimental details: the samples were fabricated using molecular beam epitaxy (MBE), we analyzed the band structure and energy levels using photoluminescence (PL) and external quantum efficiency (EQE) measurements, and the photovoltaic properties were characterized using current--voltage (\textit{J--V}) measurements. The EQE spectra and the \textit{J--V} curves were measured under single- and two-color excitation conditions. In Section~\ref{sec:Results}, we first provide the experimental results, and then we provide rate equations that are able to reproduce the experimental data in order to clarify the contribution of intraband absorption to the output current. The fitting results can explain the experimentally obtained trends well, which supports our claim of a more efficient intraband absorption and the physical picture of the problem addressed in this work. The conclusions are provided in Section~\ref{sec:Conclusions}.

\section{Experimental}\label{sec:Experimental}
\subsection{The effects of the double-tunnel-junction structure}\label{sec:Effect of doubule-tunnel-junction strucuture}
\begin{figure}[tpb]
    \includegraphics[width=\linewidth]{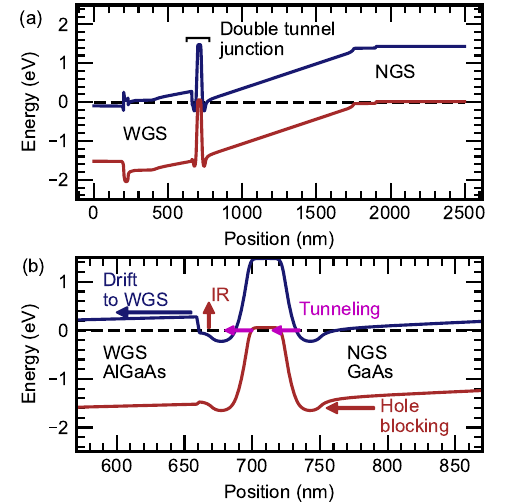}
    \caption{(a) Band diagram of the TPU-SC with a double-tunnel-junction structure (but without InAs QDs) under thermal equilibrium. The position “0 nm” corresponds to the front surface of the device. In the real device, an InAs/GaAs QD layer is located immediately below the WGS. (b) The same data as in (a) but for a narrower region.}
    \label{fig:band}
\end{figure}

Figure~\ref{fig:band}(a) shows the band diagram of our TPU-SC with a double-tunnel-junction structure under thermal equilibrium, and a magnified view of the region around the double-tunnel-junction structure is shown in Fig.~\ref{fig:band}(b). This band diagram was calculated using the semiconductor-device simulation software COMSOL Multiphysics®. In this calculation, we omitted the InAs QD layer (located at a distance of 660 nm from the front surface, immediately below the WGS layer) for simplification. The double-tunnel-junction structure is an \textit{n--p--n} structure with high doping levels and is located slightly below the position of the QD layer (in the actual device, these two structures are separated by an i-GaAs spacer layer). 

We expect three effects due to the insertion of a double-tunnel-junction structure: Firstly, the electrons in the NGS CB can reach the NGS/WGS heterointerface by band-to-band tunneling to accumulate there. Secondly, when an electron at the heterointerface absorbs an IR photon, the wavefunction of the hot electron “feels” the barrier due to the $p^+$-layer. The hot electrons generated by intraband absorption should thus have an enhanced probability to contribute to photocurrent generation by drifting to the top electrode after relaxation to the WGS CB edge, instead of diffusing to the NGS side. As a result, we expect an increase in the output current and a boost in the voltage. Thirdly, the carrier separation at the QD layer should improve, because the double-tunnel-junction structure also provides a barrier for the holes generated in the NGS layer as shown in the figure. This reduces the probability of electron--hole recombination in the QDs and improves the lifetime of electrons at the heterointerface.

\subsection{Sample fabrication}\label{sec:Sample fabrication}
We fabricated one conventional $\mathrm{Al_{0.3}Ga_{0.7}As}$/GaAs TPU-SC without tunnel junctions for reference (hereafter referred to as Device A) and one TPU-SC with a double-tunnel-junction structure (Device B) to clarify the impact of the more complex device structure. Both TPU-SCs also contained a QD layer, and the details of the two device structures are summarized in Table~\ref{tab:Structure}. The overall diode structure of Device B is \textit{n--i--n--p--n--i--p}, and the $\mathrm{Al_{0.3}Ga_{0.7}As}$/GaAs heterointerface was formed in the $i$-region.

The TPU-SCs were fabricated on $p^+$-GaAs(001) substrates by MBE. First, a 150-nm-thick \textit{p}-GaAs (Be: $2.0 \times 10^{18} /\mathrm{cm}^3$) layer followed by a 1000-nm-thick \textit{i}-GaAs layer was grown at a substrate temperature of 550 $^{\circ}$C on top of a 600 nm-thick $p^+$-GaAs (Be: $1.0 \times 10^{19} /\mathrm{cm}^3$) buffer layer. In the case of Device A, the QD layer was directly grown on top of the \textit{i}-GaAs layer. In the case of Device B, a 26-nm-thick $n^+$-GaAs (Si: $1.0 \times 10^{19} /\mathrm{cm}^3$) layer, a 28-nm-thick $p^+$-GaAs (Be: $3.0 \times 10^{19} /\mathrm{cm}^3$) layer, and a 26-nm-thick $n^+$-GaAs (Si: $1.0 \times 10^{19} /\mathrm{cm}^3$) layer for the double-tunnel-junction structure were grown first. The first two layers of the double-tunnel-junction structure were grown at a substrate temperature of 550 $^{\circ}$C, which was monitored by a pyrometer, and the third layer was grown at a lower substrate temperature of 480 $^{\circ}$C to suppress mutual diffusion of the dopants (Si and Be) and to form the abrupt band structure necessary for the tunnel junction \cite{Sugiura1988,Kojima1998}. After the growth of these three layers, the substrate temperature was raised to 585 $^{\circ}$C and kept there for 10 minutes to improve the crystallinity. We chose an overall thickness of 80 nm for the \textit{n--p--n} structure by considering the optical transparency and the tunneling probability of the tunnel junctions. Then, the 5-nm-thick \textit{i}-GaAs spacer layer was grown.

The InAs QD layer with a nominal thickness of 2.08 monolayer (ML) and a 6-nm-thick i-GaAs capping layer were grown at a substrate temperature of 490 $^{\circ}$C. The used InAs deposition amount is sufficient to induce QD formation through the Stranski--Krastanov growth mode, and the formed QDs with a density around $1.0 \times 10^{10}$ /cm$^2$ have a typical height and width of 3 nm and 20 nm, respectively \cite{Takahashi2013}. Then, the substrate temperature was elevated to 520 $^{\circ}$C to grow a 250-nm-thick \textit{i}-$\mathrm{Al_{0.3}Ga_{0.7}As}$ layer, a 150-nm-thick n-Al0.3Ga0.7As (Si: $1.0 \times 10^{17} /\mathrm{cm}^3$) layer, a 30-nm-thick $n^+$-Al$_{0.3}$Ga$_{0.7}$As (Si: $2.5 \times 10^{17}$ /cm$^3$) layer, a 30-nm-thick $n^+$-Al$_{0.8}$Ga$_{0.2}$As (Si: $2.5 \times 10^{18} /\mathrm{cm}^3$) window layer, and a 200-nm-thick $n^+$-GaAs (Si: $2.5 \times 10^{18} /\mathrm{cm}^3$) contact layer. During the MBE growth, we monitored the crystal ordering by reﬂection high energy electron diffraction (RHEED). For the electrode materials at the front and back of the SC, we used Au/AuGe and Au/AuZn, respectively.

\begin{table*}
    \vspace{-2mm}
    \centering
    \caption{The device structures of the reference TPU-SC (Device A) and the TPU-SC with a double-tunnel-junction structure (Device B) fabricated by MBE.}
    \label{tab:Structure}
    \setlength{\tabcolsep}{9pt}
    \begin{tabular}{ccccc}
        \toprule
         & Device B& & Device A & \\
        \cline{2-5}
         & Thickness & Doping concentration& Thickness & Doping concentration \\
         & (nm) & (/cm$^3$) & (nm) & (/cm$^3$) \\
        \midrule
        $n^+$-GaAs & $200$ & $2.5\times 10^{18}$ & $200$ &  $2.5\times 10^{18}$  \\
        $n^+$-Al$_{0.8}$Ga$_{0.2}$As & $30$ & $2.5\times 10^{18}$ & $30$ &  $2.5\times 10^{18}$  \\
        $n^+$-Al$_{0.3}$Ga$_{0.7}$As & $30$ & $2.5\times 10^{17}$ & $30$ &  $2.5\times 10^{17}$  \\
        $n$-Al$_{0.3}$Ga$_{0.7}$As & $150$ & $1.0\times 10^{17}$ & $150$ &  $1.0\times 10^{17}$  \\
        $i$-Al$_{0.3}$Ga$_{0.7}$As & $250$ & -- & $250$ &  --  \\
        InAs QD layer & \multicolumn{4}{c}{InAs: 0.64 nm (2.08 ML), GaAs capping-layer thickness: 6 nm}\\
        $i$-GaAs & $5$ & -- & -- &  --  \\
        $n^+$-GaAs & $26$ & $1.0\times 10^{19}$  & -- &  --  \\
        $p^+$-GaAs & $28$ & $3.0\times 10^{19}$  & -- &  --  \\
        $n^+$-GaAs & $26$ & $1.0\times 10^{19}$  & -- &  --  \\
        $i$-GaAs & $1000$ & --  & $1000$ &  --  \\
        $p$-GaAs & $150$ & $2.0\times 10^{18}$  & $150$ & $2.0\times 10^{18}$  \\
        $p^+$-GaAs & $600$ & $1.0\times 10^{19}$  & $600$ & $1.0\times 10^{19}$  \\
        \bottomrule
    \end{tabular}
\end{table*}

\subsection{EQE measurements}\label{sec:EQE measurements}
To elucidate the third effect of the double-tunnel-junction structure described in Section~\ref{sec:Effect of doubule-tunnel-junction strucuture} (\textit{i.e.}, less electron--hole recombination at the heterointerface), we measured the PL spectra of the TPU-SCs using different excitation wavelengths. The excitation power density of the used continuous-wave (CW) laser with a wavelength of 660 nm was 510 mW/cm$^2$ and that with a wavelength of 784 nm was 430 mW/cm$^2$. This corresponds to an incident photon flux density of $1.7 \times 10^{18}$ photons/(s$\cdot$cm$^2$) for both wavelengths. According to the relationship between the absorption coefficient and the excitation wavelength, excess carriers are generated in both Al$_{0.3}$Ga$_{0.7}$As and GaAs, if the 660-nm CW laser is used, whereas excess carriers are only generated in GaAs, if the 784-nm CW laser is used.

\subsection{PL measurements}\label{sec:PL measurements}
We investigated the efficiency of photocurrent generation in each device by measuring the EQE spectrum under short-circuit condition at room temperature. The EQE spectrum was measured using light from a tungsten-halogen lamp and a 140-mm single monochromator. The excitation light was modulated at a frequency of 400 Hz by an optical chopper, and the photocurrent was detected using a transimpedance amplifier and a lock-in amplifier synchronized with the optical chopper. The intensity of the monochromatic excitation light varied with the wavelength (we confirmed variation in the range of 17--32 $\upmu\mathrm{W / cm^2}$). The total integrated power density of the tungsten-halogen lamp in the wavelength range 640--1100 nm was approximately 4.7 mW/cm$^2$. We also measured the increase in the EQE that occurs if the device is additionally illuminated with IR light (from a CW laser with a wavelength of 1319 nm). The IR-beam diameter was 1.2 mm, and the excitation power density was 107 mW/cm$^2$. It is important to note that the IR photons cannot induce interband transitions in the GaAs absorber layer, since GaAs has a bandgap energy corresponding to about 870 nm. The absolute EQE value for a certain excitation condition (either single- or two-color excitation) is denoted by $\eta_{Q}$, while the increase in the EQE due to the additional IR illumination in the case of two-color excitation is denoted by $\Delta \eta_{Q}$.

\subsection{Photocurrent measurements}\label{sec:Photocurrent measurements}
We measured the short-circuit current as functions of the interband and intraband excitation power densities at room temperature, which helps us to elucidate the effects of the double-tunnel-junction structure in a TPU-SC, especially the effect concerning electron extraction after intraband absorption. For this experiment, we used a CW laser with a wavelength of 784 nm for direct band-to-band excitation of GaAs and the IR laser for intraband excitation. Both lasers were implemented in the measurement system with adjustable neutral density filters to control the excitation power densities of the two excitation beams. The excitation power density of the 784-nm light is denoted by $P_{784}$, and that of the 1319-nm light is denoted by $P_{1319}$.  The short-circuit current density obtained under two-color excitation conditions is denoted by $J_{\mathrm{SC}}$, and the short-circuit current densities obtained under single-color excitation conditions using either 784-nm light or 1319-nm light are denoted by $J_{\mathrm{SC}\_784}$ and $J_{\mathrm{SC}\_1319}$, respectively. The increase in the short-circuit density due to two-color excitation, $\Delta J_{\mathrm{SC}}$, is defined as follows:
\begin{equation}\label{eq:J_SC}
    \Delta J_{\mathrm{SC}} = J_{\mathrm{SC}} - J_{\mathrm{SC}\_784} - J_{\mathrm{SC}\_1319}.
\end{equation}
We used a source measure unit to detect the short-circuit current. The \textit{J}--\textit{V} curves of the two devices under 1 sun illumination are provided in Appendix~\ref{app:1sun}.

\section{Results and discussion}\label{sec:Results}
\subsection{PL spectra}\label{sec:PL}
\begin{figure}[tpb]
    \includegraphics[width=\linewidth]{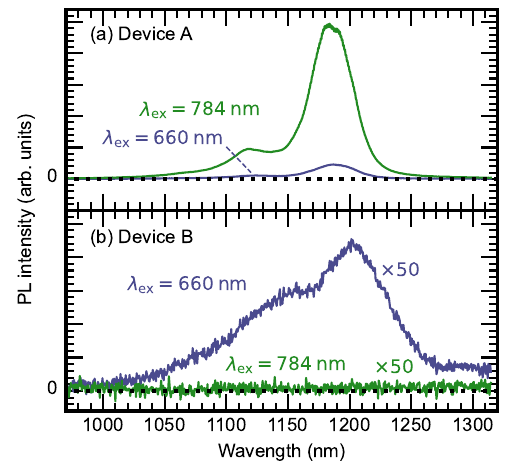}
    \caption{Room-temperature PL spectra of (a) the reference TPU-SC and (b) the TPU-SC with a double-tunnel-junction structure for 660-nm (purple) and 784-nm excitation (green).}
    \label{fig:PL}
\end{figure}

Figure \ref{fig:PL}(a) shows the room-temperature PL spectra of Device A for 660-nm and 784-nm excitation. Independent of the excitation wavelength, a relatively large PL peak appears at around 1190 nm and a smaller peak appears at around 1120 nm. These peaks are attributed to the emission from the ground and first excited states of the InAs QDs, respectively, and are consistent with our previous results \cite{Asahi2017,Kinugawa2019}. The PL spectrum obtained using 784-nm light is stronger than that obtained using 660-nm light due to the different carrier distribution: In the case of 784-nm excitation, the excitation photons pass through the Al$_{0.3}$Ga$_{0.7}$As layer and are strongly absorbed in the GaAs layer. The photogenerated electrons in GaAs drift toward the front surface due to the built-in electric field and reach the heterointerface. While some electrons are activated by thermal and tunneling processes, a large fraction of electrons will accumulate at the heterointerface. This electron accumulation partially cancels the built-in electric field, facilitating hole diffusion to the heterointerface. These holes reach the InAs QDs at the heterointerface, and thus a high recombination rate is possible. In the case of 660-nm excitation, a large fraction of photons is absorbed in the Al$_{0.3}$Ga$_{0.7}$As layer. The electrons and holes generated in the Al$_{0.3}$Ga$_{0.7}$As layer mainly recombine in the Al$_{0.3}$Ga$_{0.7}$As layer. Hence, the PL emission is weaker.

The PL spectra of Device B under the same excitation conditions are shown in Fig.~\ref{fig:PL}(b). Compared to Device A, the PL intensities of Device B are substantially weaker. In the case of 660-nm excitation, the majority of carriers recombines in the Al$_{0.3}$Ga$_{0.7}$As as explained above, resulting in a relatively weak PL emission. Note that the PL spectrum under 660-nm excitation is weaker than that in Fig.~\ref{fig:PL}(a). This is because, due to the insertion of the double-tunnel junction, photogenerated carriers in the GaAs layer do not contribute to the PL emission from the QDs. In the case of 784-nm excitation, no PL emission is observed. This is one of the expected effects of the double-tunnel-junction structure: a suppression of electron--hole recombination in the InAs QDs, since almost no excess holes in GaAs reach the heterointerface. The PL due to absorption of photons in the GaAs capping layer of the QDs and the spacer layer was not observed, because the 11-nm-thick GaAs capping and spacer layers are insufficient to generate enough carriers to to produce a detectable PL signal with our facilities.
\subsection{EQE spectra}\label{sec:EQE}
\begin{figure}[tpb]
    \includegraphics[width=\linewidth]{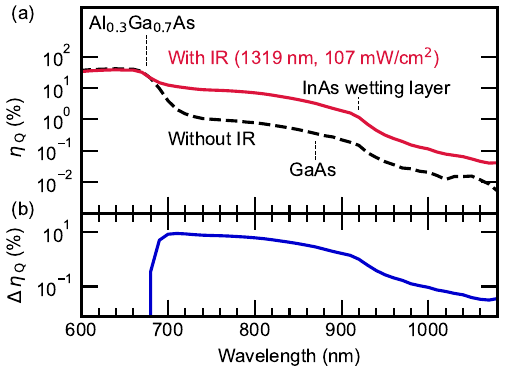}
    \caption{(a) The room-temperature EQE spectra of the TPU-SC with a double-tunnel junction for single- (black dashed curve) and two-color excitation (red solid curve). (b) The IR-induced change in the EQE, obtained by subtracting the EQE measured without the IR light from that measured with additional IR irradiation.}
    \label{fig:EQE}
\end{figure}

Figure \ref{fig:EQE}(a) shows the EQE spectra of Device B. The black dashed curve represents the results obtained without additional IR irradiation. A distinct EQE drop appears at a wavelength of about 690 nm, which corresponds to the absorption edge of Al$_{0.3}$Ga$_{0.7}$As. Another slight drop appears at 920 nm, which corresponds to the energy level of the InAs wetting layer. Note that no clear absorption edge appears at the wavelength corresponding to the bandgap of GaAs due to the \textit{n--p--n} double-tunnel-junction structure. The red solid curve represents the EQE spectrum obtained with additional IR irradiation. The IR photons mainly excite the electrons accumulated at the heterointerface. Especially in the wavelength region 690--870 nm, the electron accumulation at the heterointerface is significant, and thus the IR-induced EQE increase, $\Delta \eta_{Q}$, reaches its maximum in this wavelength region as shown in Fig.~\ref{fig:EQE}(b). Regarding the $\Delta \eta_{Q}$ in the wavelength region > 870 nm, the values in Fig. \ref{fig:EQE}(b) are higher than those of the TPU-SC without a double-tunnel-junction structure shown in Fig. 2(b) in Ref. \cite{Watanabe2021} (\textit{e.g.}, $\approx 0.27$\% vs. $\approx 0.01$\% at 950 nm). The sub-bandgap photons from the tungsten-halogen lamp induce interband transitions in the tunnel junction and lead to an accumulation of electrons at the heterointerface, resulting in a clearly measurable $\Delta \eta_{Q}$ even in this wavelength range.
\subsection{Interband excitation power dependence of short-circuit current}\label{sec:PowdepJSC}
\begin{figure*}[tb]
    \includegraphics[width=\linewidth]{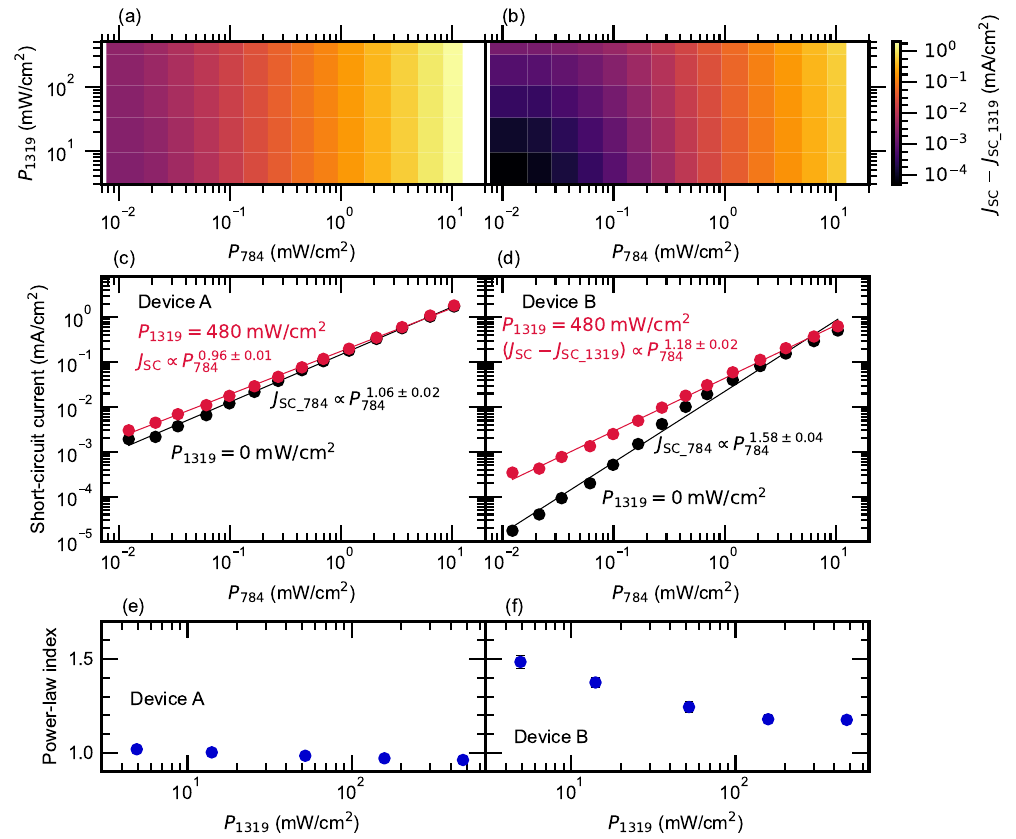}
    \caption{The short-circuit current density as functions of the excitation power intensities of the two excitation beams for (a) Device A and (b) Device B. The horizontal axis indicates the intensity of the 784-nm excitation beam ($P_{784}$) and the vertical axis represents the intensity of 1319-nm excitation beam ($P_{1319}$). Note that we plotted $J_{\mathrm{SC}} - J_{\mathrm{SC}\_1319}$ as functions of $P_{784}$ and $P_{1319}$, where $J_{\mathrm{SC}\_1319}$ denotes the corresponding short-circuit current density obtained under single-color excitation conditions using 1319-nm light. (c) The dependence of the short-circuit current density on $P_{784}$ for $P_{1319}=0$ (black closed circles) and 480 mW/cm$^2$ (red closed circles) in the case of Device A, and (d) the data obtained from Device B. The black and red lines indicate the results of fitting the relationship $J_\mathrm{SC} \propto P_{784}^n$ to the different data sets, where n is the power-law index. (e) The obtained power-law index as a function of $P_{1319}$ in the case of Device A, and (f) the data obtained from Device B. The error bars represent the standard error for the obtained regression lines.}
    \label{fig:J_SC}
\end{figure*}

The two-dimensional map in Fig.~\ref{fig:J_SC}(a) shows the measured short-circuit current density as functions of the excitation power densities of the two excitation beams ($P_{784}$ and $P_{1319}$) for Device A, and Fig.~\ref{fig:J_SC}(b) shows that for Device B. To remove the effects of IR-induced interband transitions observed in the case of Device B, we subtracted the corresponding short-circuit current density values observed for single-color excitation using 1319-nm light, $J_{\mathrm{SC}\_1319}$.  In the case of Device A, $J_{\mathrm{SC}\_1319}$ is negligible, even under strong excitation conditions. The details of the impact of single-color excitation using IR light are shown in Appendix~\ref{app:Single_ex} and the raw data for Fig.~\ref{fig:J_SC}(b) (before subtraction) is shown in Appendix~\ref{app:J_SC_debB}. In Figs.~\ref{fig:J_SC}(c) and \ref{fig:J_SC}(d), we provide the data sets obtained using $P_{1319} = 0$ and 480 $\mathrm{mW / cm^2}$ for each sample. From these data it is evident that the output current increases with both $P_{784}$ and $P_{1319}$. A higher $P_{784}$ leads to a higher electron density at the heterointerface, and thus improves the occupation probability of the initial state in the second step of the TPU process. This explains the observed increasing trend of the output current.

We examined the characteristics of the data in Figs.~\ref{fig:J_SC}(c) and \ref{fig:J_SC}(d) using a power-law function, \textit{i.e.}, we fitted the following equation to each data set:
\begin{equation}\label{eq:propt_J_SC}
    J_{\mathrm{SC}\_784} \propto P_{784}^{n},
\end{equation}
where $n$ is the power-law index (the results are shown as solid curves). The data of Device A can be well approximated by a linear function, in agreement with our previous results \cite{Asahi2017}. The slightly superlinear behavior described by the fitting result $n = 1.06\;\pm0.02$ for single-color excitation is probably caused by the fact that the 784-nm light simultaneously induces band-to-band transitions as well as intraband transitions at the heterointerface.

The red closed circles in Fig.~\ref{fig:J_SC}(c) are the short-circuit current data obtained using two-color excitation, $J_{\mathrm{SC}}$, where 1319-nm light is used to selectively excite the electrons accumulated at the heterointerface. The $J_{\mathrm{SC}}$ values for two two-color excitation are larger than those obtained without the 1319-nm light, demonstrating the effect of IR-induced intraband transitions at the heterointerface. The obtained $n$ in the case of $P_{1319} = 480\; \mathrm{mW / cm^2}$ is 0.96 $\pm$0.01, which means that the power-law index decreased by applying the 1319-nm light and the device response became even slightly sublinear. Figure~\ref{fig:J_SC}(e) summarizes the obtained power-law index of Device A for different values of $P_{1319}$: $n$ decreases as $P_{1319}$ is increased. We interpret this tendency as follows: At low values of $P_{784}$, the 1319-nm photons govern the intraband absorption, but the initially dominant role of the 1319-nm photons is taken over by the 784-nm photons when $P_{784}$ is increased. This results in a lower power-law index compared to that observed without IR irradiation, and since the impact of the 1319-nm photons can be observed up to higher values of $P_{784}$ if $P_{1319}$ is larger, n decreases as $P_{1319}$ is increased. 

Figures~\ref{fig:J_SC}(d) and (f) clarify that the data of Device B exhibit larger power-law indices. The data for single-color excitation in Fig.~\ref{fig:J_SC}(d) (the black closed circles) follow a power law with an exponent of 1.58 $\pm$0.01. We consider that the observed stronger superlinear relationship is mainly caused by the effects of the double-tunnel-junction structure described in Section~\ref{sec:Effect of doubule-tunnel-junction strucuture}. In the case of Device A, a significant number of CB electrons excited by 784-nm photons at the heterointerface diffuses to the GaAs side, resulting in a low photocurrent extraction efficiency. On the other hand, the barrier layer of the double-tunnel-junction structure effectively suppresses diffusion towards the GaAs side, and thus current generation through intraband absorption of 784-nm photons at the heterointerface can occur more efficiently in Device B. Furthermore, the data obtained with additional 1319-nm irradiation [Fig.~\ref{fig:J_SC}(d); red closed circles] show a significant increase in the short-circuit current density. However, compared to the results of Device A, the current density levels are lower, which is primarily due to a lower rate of thermal activation of electrons at the CB offset in Device B: Due to the electron-energy dependence of the transmission probability of the tunnel junction, the electron energy distribution at the heterointerface deviates from the normal Fermi--Dirac distribution (\textit{i.e.}, a non-thermal electron distribution is established). In particular, the number of high-energy electrons decreases. As a result, the thermal activation of electrons over the barrier at the heterointerface is suppressed, leading to a significant reduction in the short-circuit current density. This effect is particularly pronounced under weak excitation conditions, because this mechanism is usually the main cause of the output current in this regime. This suppression of the short-circuit current generated via thermal activation is partially compensated by a higher $P_{784}$ (because the contribution of intraband absorption to the output current becomes significant), eventually making the current density levels more comparable to that of Device A.

We would like to point out that the power-law index of 1.18 $\pm$0.02 for $P_{1319} = 480\: \mathrm{mW / cm^2}$ in the case of Device B is larger than that of Device A, which indicates that the intraband transitions caused by 784-nm light in Device B have a larger contribution than in Device A. 
\clearpage

\subsection{Interband excitation power dependence of change in the short-circuit current}\label{sec:Powdep_Delta_JSC}
\begin{figure*}[tb]
    \includegraphics[width=\linewidth]{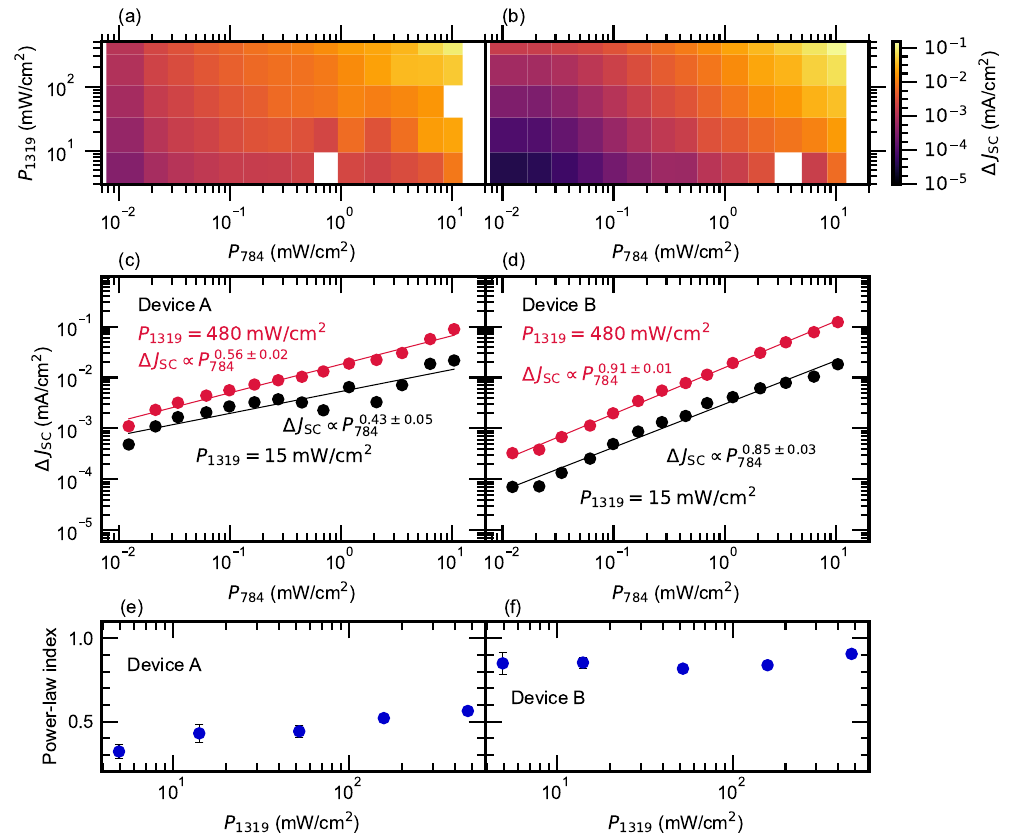}
    \caption{The two-dimensional map of $\Delta J_\mathrm{SC}$ as functions of $P_{784}$ (horizontal axis) and $P_{1319}$ (vertical axis) for (a) Device A and (b) Device B. (c) The dependence of $\Delta J_\mathrm{SC}$ on $P_{784}$ for $P_{1319}=$ 15 mW/cm$^2$ (black closed circles) and 480 mW/cm$^2$ (red closed circles) and the corresponding fitting results (solid lines) in the case of Device A. (d) The experimental results and fitting results in the case of Device B. (e) The dependence of the power-law index on $P_{1319}$ in the case of Device A, and (f) the data obtained from Device B. The error bars represent the standard error for the obtained regression lines.}
    \label{fig:delta_J_SC}
\end{figure*}

Figures~\ref{fig:delta_J_SC}(a) and (b) present the two-dimensional maps of the $\Delta J_{\mathrm{SC}}$ values of Device A and Device B under two-color excitation conditions, respectively, calculated using Eq.~(\ref{eq:J_SC}). $\Delta J_{\mathrm{SC}}$ increases with both $P_{784}$ and $P_{1319}$ due to the same reasons as explained for the two-color-excitation data in Fig.~\ref{fig:J_SC}. Figures~\ref{fig:delta_J_SC}(c) and (d) show cross-sectional profiles of the maps in Figs.~\ref{fig:delta_J_SC}(a) and (b), respectively. Similar to the above discussion of $J_{SC}$, we fitted a power-law function to the data:
\begin{equation}\label{eq:propt_Delta_J_SC}
    \Delta J_\mathrm{SC} \propto P_{784}^{n}.
\end{equation}
The fitting results for $P_{1319} = 15\: \mathrm{mW / cm^2}$ and $480\: \mathrm{mW / cm^2}$ are shown by the black and red lines, respectively. In Device A, the power-law index $n$ of  $\Delta J_{\mathrm{SC}}$  is well below unity. For $P_{1319} = 15\: \mathrm{mW / cm^2}$, $n = 0.43 \pm0.05$. Because the probability of intraband absorption is proportional to the electron density of the initial state, we believe that the observed strongly sublinear tendency of $\Delta J_{\mathrm{SC}}$ is caused by the sublinear increase of the electron density at the heterointerface with $P_{784}$ as explained above (the stronger electron accumulation at the heterointerface at higher power densities weakens the electric field and facilitates electron--hole recombination). For $P_{1319} = 480\: \mathrm{mW / cm^2}$, we obtained an $n$ of 0.56 $\pm$0.02, and Fig.~\ref{fig:delta_J_SC}(e) summarizes dependence of $n$ on $P_{1319}$. The power-law index gradually increases with the IR excitation power density, which is explained as follows: When the intraband absorption rate increases, the contribution of electron--hole recombination becomes less effective, resulting in a higher power-law index.

In contrast to these rather low $n$ values, the $\Delta J_{\mathrm{SC}}$ data of Device B in Fig.~\ref{fig:delta_J_SC}(d) follow power-law functions with much larger exponents of 0.85 $\pm$0.03 and 0.91 $\pm$0.01. This is attributed to the better carrier separation due to the double-tunnel-junction structure. In particular, the $n^+$-layer of the double-tunnel-junction structure functions as a barrier layer that suppresses holes in the GaAs layer from reaching the heterointerface. Consequently, the recombination probability at the heterointerface decreases, leading to a relatively linear increase in the output current with $P_{784}$. These results indicate that the double-tunnel-junction structure can lead to a higher current generation efficiency through intraband absorption. The obtained $n$ values are slightly less than unity due to a significant contribution of 784-nm photons to the overall intraband transition rate: With increasing $P_{784}$, the role of the 1319-nm photons for intraband absorption is taken over by the 784-nm photons, which causes an underestimation of the “true” $n$ value related to TPU resulting in values slightly less than unity. Figure~\ref{fig:delta_J_SC}(f) summarizes the dependence of the power-law index of $\Delta J_{\mathrm{SC}}$ on $P_{1319}$. The data suggest that the power-law index is basically independent of $P_{1319}$ as a result of a relatively large contribution of intraband absorption of IR photons to the output current even under weak IR irradiation.
\clearpage
\subsection{Simulation using rate equations}\label{sec:rate_equations}
In the previous sections, we discussed the experimentally observed characteristics related to intraband transitions. Because the 784-nm photons induce both interband and intraband transitions and the contribution of the 784-nm photons to the overall intraband transition rate is altered by the 1319-nm light, it is difficult to experimentally isolate the pure contribution of the 1319-nm light to the output current. Therefore, we established a rate-equation system that is able to reproduce the observed $J_{\mathrm{SC}}$  trends. This method of analyzing the current generation mechanism in SCs allows us to obtain important physical parameters of the investigated SC by a fitting procedure \cite{Tamaki2015,Asahi2016}. While rate equations are too simple to fully reproduce all phenomena occurring in a SC, we believe that they can help us to identify the essential impact of a given excitation beam on the carrier dynamics, while avoiding the need of performing more complex simulations.

\subsubsection{The rate-equation system}\label{subsec:rate_equations}
\begin{figure}[tpb]
    \includegraphics[width=\linewidth]{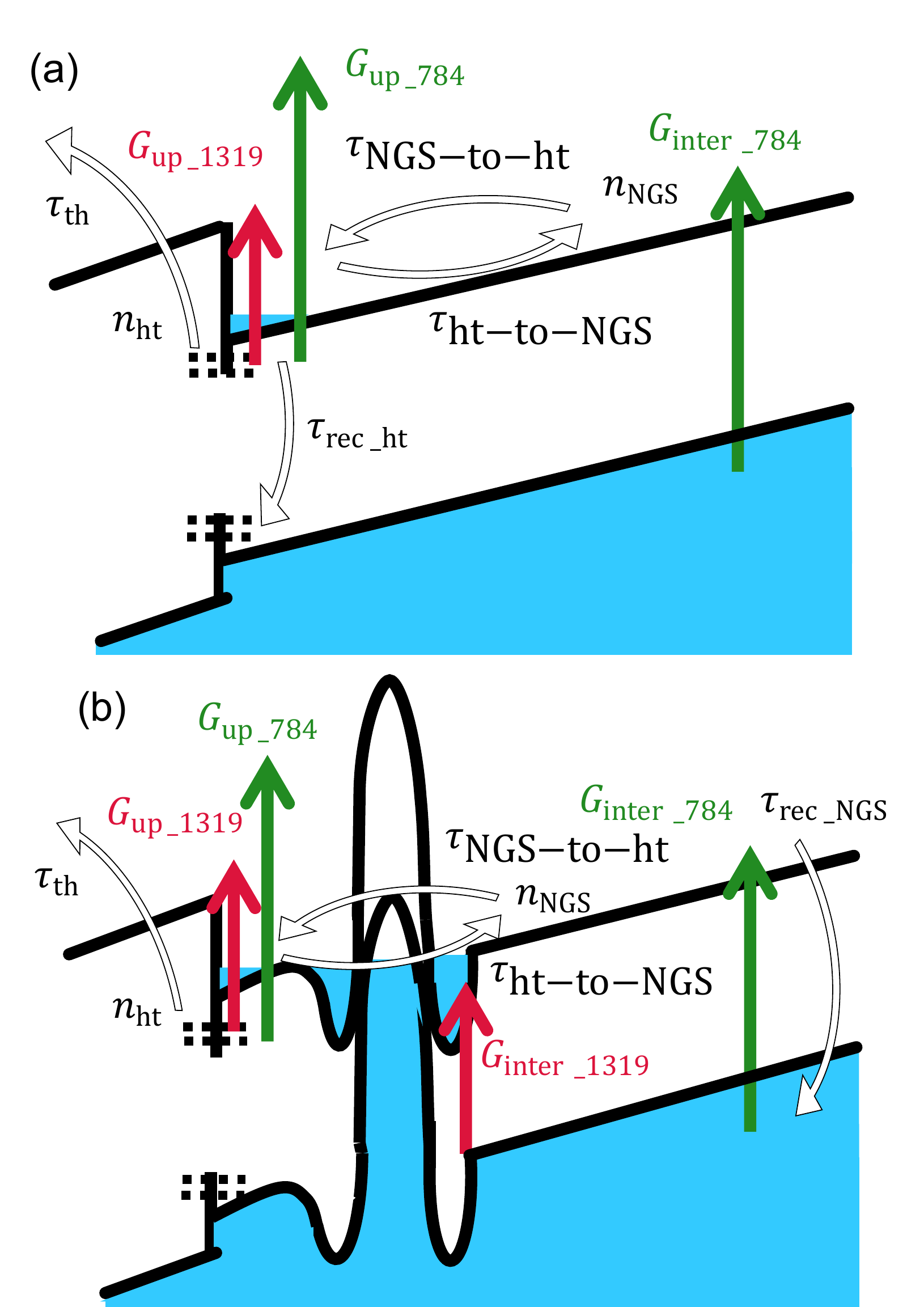}
   \caption{Calculation model for (a) Device A and (b) Device B.}
    \label{fig:model}
\end{figure}

The transitions considered in our rate-equation system in the case of Device A are summarized in Fig.~\ref{fig:model}(a). The green arrow on the right-hand side, $G_\mathrm{inter\_784}$, is the generation rate of CB electrons in the NGS layer by 784-nm light. The excess carrier density in the NGS CB is denoted by $n_\mathrm{NGS}$. This carrier reservoir is connected to the electron density at the heterointerface, $n_\mathrm{ht}$, where several processes need to be considered:
\begin{equation}\label{eq:rate_1}
    \frac{\partial n_\mathrm{NGS}}{\partial t} = G_\mathrm{inter\_784} - \frac{n_\mathrm{NGS}}{\tau_\mathrm{NGS\mathchar`-to\mathchar`-ht}} + \frac{n_\mathrm{ht}}{\tau_\mathrm{ht\mathchar`-to\mathchar`-NGS}},
\end{equation}
\begin{equation}\label{eq:rate_2}
    \begin{split}
        \frac{\partial n_\mathrm{ht}}{\partial t} &= \frac{n_\mathrm{NGS}}{\tau_\mathrm{NGS\mathchar`-to\mathchar`-ht}} - \frac{n_\mathrm{ht}}{\tau_\mathrm{ht\mathchar`-to\mathchar`-NGS}} - \frac{n_\mathrm{ht}}{\tau_\mathrm{rec\_ht}} \\ &\quad + G_\mathrm{up\_784} - G_\mathrm{up\_1319} - \frac{n_\mathrm{ht}}{\tau_\mathrm{th}}.
    \end{split}
\end{equation}
Here, $\tau_\mathrm{NGS\mathchar`-to\mathchar`-ht}$ and $\tau_\mathrm{ht\mathchar`-to\mathchar`-NGS}$ are the time constants for electron transfer from the NGS CB to the heterointerface and vice versa, respectively, $\tau_\mathrm{rec\_ht}$  is the electron lifetime at the heterointerface, $G_\mathrm{up\_784}$ and $G_\mathrm{up\_1319}$  are the intraband transition rates due to 784-nm and the 1319-nm photons, respectively, and $\tau_\mathrm{th}$ is the time constant for the thermal activation of electrons at the heterointerface. Note that the electron density considered here is the areal electron density in units of electrons/cm$^{-2}$. In Eqs.~(\ref{eq:rate_1}) and (\ref{eq:rate_2}), we omitted the electron--hole recombination in the NGS layer, because the experimentally observed PL intensity from GaAs is weak compared to that from the InAs QDs. Furthermore, we assumed that the electron--hole recombination rate at the heterointerface is independent of the hole density and that the time constants used in Eqs.~(\ref{eq:rate_1}) and (\ref{eq:rate_2}) are independent of the excitation conditions.

We established a similar rate-equation system for Device B:
\begin{equation}\label{eq:rate_3}
    \begin{split}
        \frac{\partial n_\mathrm{NGS}}{\partial t} &= G_\mathrm{inter\_784} + G_\mathrm{inter\_1319} \\
        &\quad - \frac{n_\mathrm{NGS}}{\tau_\mathrm{NGS\mathchar`-to\mathchar`-ht}} + \frac{n_\mathrm{ht}}{\tau_\mathrm{ht\mathchar`-to\mathchar`-NGS}} + \frac{n_\mathrm{NGS}}{\tau_\mathrm{rec\_NGS}},
    \end{split}
\end{equation}
\begin{equation}\label{eq:rate_4}
    \begin{split}
        \frac{\partial n_\mathrm{ht}}{\partial t} &= \frac{n_\mathrm{NGS}}{\tau_\mathrm{NGS\mathchar`-to\mathchar`-ht}} - \frac{n_\mathrm{ht}}{\tau_\mathrm{ht\mathchar`-to\mathchar`-NGS}} \\
        &\quad + G_\mathrm{up\_784} - G_\mathrm{up\_1319} - \frac{n_\mathrm{ht}}{\tau_\mathrm{th}},
    \end{split}
\end{equation}
and the transitions considered here are summarized in Fig.~\ref{fig:model}(b). Since the holes in the NGS VB hardly reach the heterointerface, we added a term for the electron--hole recombination in the NGS layer with the time constant $\tau_\mathrm{rec\_NGS}$  and removed the corresponding term from the equation for the heterointerface. Furthermore, as shown in Appendix~\ref{app:Single_ex}, the 1319-nm photons also cause interband transitions in the case of Device B. Therefore, we added the IR-related interband-excitation term $G_\mathrm{inter\_1319}$ to Eq.~(\ref{eq:rate_3}).

Based on the Beer--Lambert law $A=1-\exp (-\alpha d)$, where $A$ is the absorptivity, $\alpha$ is the absorption coefficient, and $d$ is absorber thickness, we consider that the intraband transition rates in Eqs.~(\ref{eq:rate_2}) and (\ref{eq:rate_4}) obey the following equations:
\begin{equation}\label{eq:rate_up}
    G_\mathrm{up\_784} = N_{784}[1 - \exp (-k_{784} n_\mathrm{ht})],
\end{equation}
\begin{equation}\label{eq:rate_up2}
    G_\mathrm{up\_1319} = N_{1319}[1 - \exp (-k_{1319} n_\mathrm{ht})],
\end{equation}
where $N_{784}$ and $N_{784}$ denote the photon fluxes of the 784-nm and 1319-nm excitation beams, respectively. As the intraband absorption coefficient is proportional to the carrier density in the initial state, it can be also written as $\alpha = \alpha_\mathrm{max}n/n_\mathrm{max}$, where $\alpha_\mathrm{max}$ is the absorption coefficient for maximum occupation of the initial state and $n_\mathrm{max}$ is the corresponding carrier density. Regarding Eqs.~(\ref{eq:rate_up}) and (\ref{eq:rate_up2}), the absorber thickness $d$, the maximum absorption coefficient  $\alpha_\mathrm{max}$, and the maximum carrier density $n_\mathrm{max}$ are unknown, and are therefore treated as a single fitting parameter: $k = (d \alpha_\mathrm{max})/n_\mathrm{max}$. Hence, $k_{784}$ and $k_{1319}$ are the effective cross-sections for intraband absorption of 784-nm and 1319-nm photons at the heterointerface, respectively.

Likewise, we used the following equations to calculate the interband transition rates in Eqs.~(\ref{eq:rate_1}) and (\ref{eq:rate_3}):
\begin{equation}\label{eq:rate_inter1}
    G_\mathrm{inter\_784} = N'_{784}[1 - \exp (-\alpha_{784} d_\mathrm{NGS})],
\end{equation}
\begin{equation}\label{eq:rate_inter2}
    G_\mathrm{inter\_1319} = N'_{1319}[1 - \exp (-\alpha_{1319} d'_\mathrm{ht})].
\end{equation}
In these equations, $N'_{784}$ and $N'_{1319}$ denote the photon fluxes of the 784-nm and 1319-nm excitation beams that reach the NGS layer and the double-tunnel-junction structure, respectively, $\alpha_{784}$ and $\alpha_{1319}$  are the absorption coefficients for these photons, $d_\mathrm{NGS}$  is the thickness of the NGS layer, and $d'$ is the effective thickness of the double-tunnel-junction structure. Since, we assume that the absorption of 784-nm photons in the highly doped layers of the double-tunnel-junction structure is negligible, the connection with Eqs.~(\ref{eq:rate_up}) and (\ref{eq:rate_up2}) is as follows:
\begin{equation}\label{eq:rate_N1}
    N'_{784} = N_{784} - G_\mathrm{up\_784},
\end{equation}
\begin{equation}\label{eq:rate_N2}
    N'_{1319} = N_{1319} - G_\mathrm{up\_1319}.
\end{equation}
Based on the used NGS and the sample structure, we used $\alpha_{784} = 14000 \;\mathrm{cm^{-1}}$ \cite{Edward1985} and $d_\mathrm{NGS} = 1.0 \;\upmu\mathrm{m}$  in our calculations. Additionally, we introduced the absorptivity for IR-induced interband transitions at the double-tunnel-junction structure, $A_\mathrm{inter\_1319}$:
\begin{equation}\label{eq:rate_A}
    A_\mathrm{inter\_1319} = 1 - \exp (- \alpha_{1319} d'),
\end{equation}
which was substituted for the term in the square brackets in Eq.~(\ref{eq:rate_inter2}) and treated as a fitting parameter.

By numerically solving the rate equations for various values of $P_{784}$ and $P_{1319}$ under steady-state conditions, we determined the values of the electron densities $n_\mathrm{NGS}$ and $n_\mathrm{ht}$  as functions of $P_{784}$ and $P_{1319}$ for both devices [the time derivatives in Eqs.~(\ref{eq:rate_1})--(\ref{eq:rate_4}) were set to zero and the Newton--Raphson method was employed]. Then, the total output current was calculated using the obtained electron density at the heterointerface as follows:
\begin{align}\label{eq:calc_J}
        J_\mathrm{SC} &= qN_{784}[1 - \exp (- k_{784} n_\mathrm{ht})] \notag \\
                    &\quad + qN_{1319}[1 - \exp (- k_{1319} n_\mathrm{ht})]  \notag \\
                    &\quad + q\frac{n_\mathrm{ht}}{\tau_\mathrm{th}}  \notag \\
                    &= J_\mathrm{up\_784} + J_\mathrm{up\_1319} + J_\mathrm{up\_th}.
\end{align}
Here, $q$ is the elementary charge, $J_\mathrm{up\_784}$ and $J_\mathrm{up\_1319}$ are the current components generated by intraband absorption of 784-nm and 1319-nm photons, respectively, and $J_\mathrm{up\_th}$ is the current component generated by thermal activation. During the fitting of Eq.~(\ref{eq:calc_J}) to the experimental $J_\mathrm{SC}$ data of Fig.~\ref{fig:J_SC}, we tried to minimize the total sum of the squares of the residuals and used $\tau_\mathrm{NGS\mathchar`-to\mathchar`-ht}$, $\tau_\mathrm{ht\mathchar`-to\mathchar`-NGS}$, $\tau_\mathrm{rec\_NGS}$, $\tau_\mathrm{rec\_ht}$, $k_{784}$, $k_{1319}$, and $A_\mathrm{inter\_1319}$ as global fitting parameters (the final values of these parameters are provided in Table~\ref{tab:sim}). Due to the large number of free parameters, it is considered inappropriate to apply the common gradient method, because this method depends on the initial values and tends to converge to local optima. To find the global optimum solution, we employed a genetic algorithm (a type of metaheuristic optimization) \cite{Katoch2021,Whitley1994}. The optimization was implemented using Pymoo \cite{Blank2020}.

\subsubsection{Simulation results}\label{subsec:results}
\begin{figure*}[tpb]
    \includegraphics[width=\linewidth]{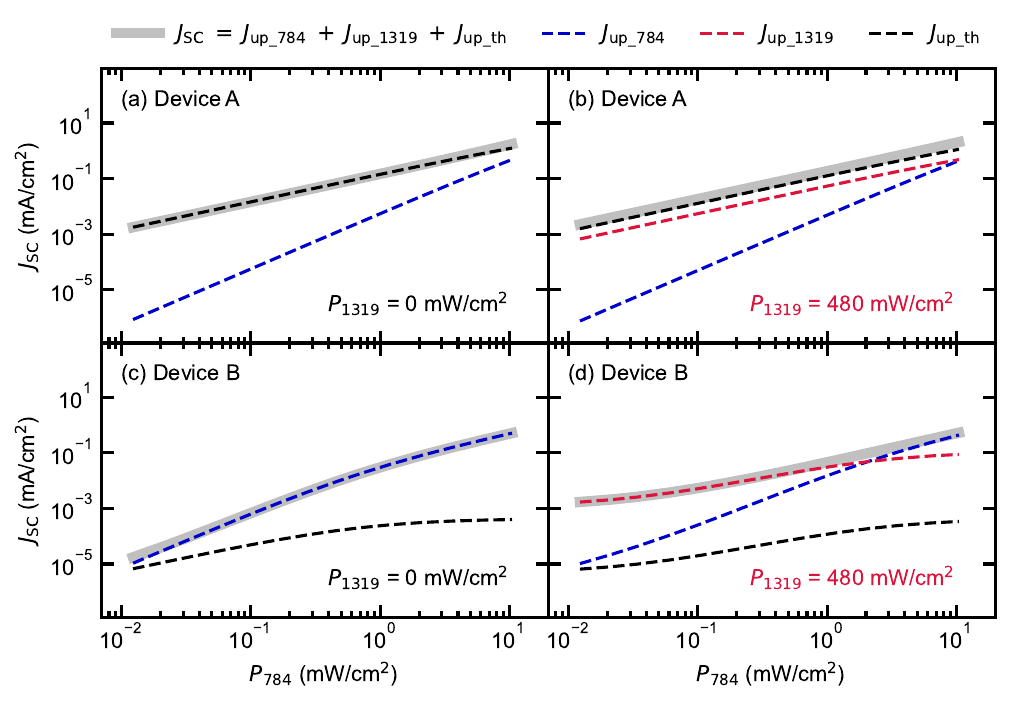}
    \caption{(a) The dependence of $J_\mathrm{SC}$ (gray curve) on $P_{784}$ and its breakdown for Device A under single- and (b) two-color excitation conditions, calculated using the rate equations with the parameters provided in Table~\ref{tab:sim}. (c) The calculation results for Device B under single- and (d) two-color excitation conditions. The blue and red dashed curves indicate the current components due to the intraband absorption of 784-nm and 1319-nm photons, respectively. The black dashed curves indicate the thermally induced component.}
    \label{fig:sims}
\end{figure*}

The comparison between the experimental and calculated data sets is shown in Appendix~\ref{app:sim_comp}, verifying that the calculated trends accurately reproduce the experimentally observed trends for a rather wide range of experimental conditions. Figure~\ref{fig:sims} presents a breakdown of the calculated $J_\mathrm{SC}$ values according to Eq.~(\ref{eq:calc_J}). Figure~\ref{fig:sims}(a) presents the results for Device A in the case of single-color excitation. The total output current is governed by the thermally induced component, which is indicated by the black dashed curve and increases almost linearly with $P_{784}$, consistent with the description in Section~\ref{sec:PowdepJSC}. As the interband absorption rate is increased, the contribution of intraband absorption of 784-nm photons to the output current (indicated by the blue dashed line) increases superlinearly and approaches the thermally induced component. Figure~\ref{fig:sims}(b) shows the data for two-color excitation using $P_{1319} = 480 \mathrm{mW/cm^2}$. Although the current generated by thermal activation remains the main component, the contribution of intraband absorption of 1319-nm photons to the output current (indicated by the red dashed curve) is significant. This is clear evidence of the photocurrent enhancements that can be achieved by TPU processes in a conventional TPU-SC with QDs \cite{Asahi2017}.

Figure~\ref{fig:sims}(c) shows the calculation results for Device B in the case of single-color excitation. In contrast to the data shown in Fig.~\ref{fig:sims}(a), the output current of Device B is dominated by the current generated through intraband absorption of 784-nm photons, and the contribution of thermal activation to the output current becomes smaller as $P_{784}$ is increased. This characteristic reflects the superlinear behavior of the short-circuit current shown in Fig.~\ref{fig:J_SC}(d). In addition, the $J_\mathrm{up\_th}$ values of Device B are much smaller than those of Device A. As described in Section~\ref{sec:PowdepJSC}, the electron energy distribution in Device B deviates from the normal Fermi--Dirac distribution (the thermal population of higher energy levels is suppressed). Hence, the probability for thermal escape is lower, which leads to a lower $J_\mathrm{up\_th}$. Figure~\ref{fig:sims}(d) shows the data in the case of two-color excitation using $P_{1319} = 480\;\mathrm{mW/cm^2}$. At lower interband excitation power densities, the current component due to the 1319-nm-induced intraband transitions (indicated by the red dashed curve) is the strongest component. As the interband excitation power density is increased, the current component due to the 784-nm-induced intraband transitions superlinearly increases and finally becomes the dominant component, indicating that the efficiency of current generation through 784-nm photons is higher than that of 1319-nm photons. These curves prove that a double-tunnel-junction structure can lead to a larger contribution of intraband absorption to the total output current.

Regarding the obtained effective cross-sections for intraband absorption shown in Table~\ref{tab:sim}, $k_{784}$ and $k_{1319}$, the values for Device B are higher than those for Device A by two to three orders of magnitude. This indicates that the insertion of a double-tunnel-junction structure improves the contribution of intraband absorption to the output current. Regarding the inverse of the thermal activation rate of the electrons at the heterointerface, $\tau_\mathrm{th}$, the value for Device B is longer than that for Device A by a factor of three. This indicates the aforementioned effect of a suppression of the electron population at higher energy levels.

Thus, the obtained parameters and results shown in Fig. 7 support our claim that the insertion of a double-tunnel-junction structure improves the efficiency of intraband absorption. The proposed structure may pave the way toward the realization of high-efficiency SCs utilizing intraband absorption of sub-bandgap photons. 

\begin{table*}[tb]
    \vspace{-5mm}
    \centering
    \caption{The parameters of the regression lines, obtained by global fitting of Eqs.~(\ref{eq:rate_1})--(\ref{eq:rate_4}) to the experimentally obtained $J_\mathrm{SC}$ data sets: $\tau_\mathrm{NGS\mathchar`-to\mathchar`-ht}$ and $\tau_\mathrm{ht\mathchar`-to\mathchar`-NGS}$ are the time constants for electron transfer from the NGS to the heterointerface and vice versa, respectively, $\tau_\mathrm{rec\_ht}$ and $\tau_\mathrm{rec\_NGS}$ are the electron lifetimes at the heterointerface and in the NGS layer, respectively, $\tau_\mathrm{th}$ is the time constant for thermal activation of electrons at the heterointerface, $k_{784}$ and $k_{1319}$ are the effective cross-sections for intraband absorption of 784-nm and 1319-nm photons, respectively, and $A_\mathrm{inter\_1319}$ denotes the interband absorptivity of 1319-nm light of the double-tunnel-junction structure.}
    \label{tab:sim}
    \setlength{\tabcolsep}{5pt}
    \begin{tabular}{cccccccc}
        \toprule
         & $\tau_\mathrm{NGS\mathchar`-to\mathchar`-ht}$ (s) & $\tau_\mathrm{ht\mathchar`-to\mathchar`-NGS}$ (s) & $\tau_\mathrm{rec\_ht}$,& $\tau_\mathrm{th}$ (s) &  $k_{784}$ (cm$^2$ &$k_{1319}$ (cm$^2$ & $A_\mathrm{inter\_1319}$ \\
         & & & $\tau_\mathrm{rec\_NGS}$ (s) & & /electron) & /electron) & \\
        \midrule
        Device A & $8.64 \times 10^{-5}$ & $7.42 \times 10^{-5}$ & $2.02 \times 10^{-8}$ &  $4.62 \times 10^{-8}$ & $2.11 \times 10^{-10}$ & $2.89 \times 10^{-12}$ & -- \\
        Device B & $6.38 \times 10^{-6}$ & $1.19 \times 10^{-9}$ & $8.76 \times 10^{-7}$ &  $1.39 \times 10^{-7}$ & $2.35 \times 10^{-7}$ & $5.87 \times 10^{-10}$ & $2.71 \times 10^{-5}$ \\
        \bottomrule
    \end{tabular}
\end{table*}
\clearpage
\section{Conclusions}\label{sec:Conclusions}
We have demonstrated that a higher intraband-absorption efficiency can be achieved in a TPU-SC by inserting a double-tunnel-junction structure. The absence of PL from the QDs at the heterointerface in the case of direct excitation of the NGS layer indicates that hole diffusion to the heterointerface is successfully suppressed by the double-tunnel-junction structure. This feature enhances the probability of intraband absorption for a given electron injection rate at the heterointerface, because the electron lifetime at the heterointerface is extended. We have shown that the $J_\mathrm{SC}$ of the TPU-SC with a double-tunnel-junction structure superlinearly increases with the primary excitation beam for direct excitation of the NGS, which indicates that this light can also induce a significant amount of intraband transitions. In addition, a simulation using rate equations indicates a higher intraband absorption efficiency in the case of the TPU-SC with a double-tunnel-junction structure.

\begin{acknowledgments}
This work was supported by the Japan Society for the Promotion of Science (JSPS) KAKENHI Grant Numbers JP23H01448 and JP23K03958, and the JST-ALCA-Next Japan Grant Number JPMJAN24B1.
\end{acknowledgments}

\appendix
\section{\textit{J}--\textit{V} curves}\label{app:1sun}
\begin{figure}[tpb]
    \includegraphics[width=\linewidth]{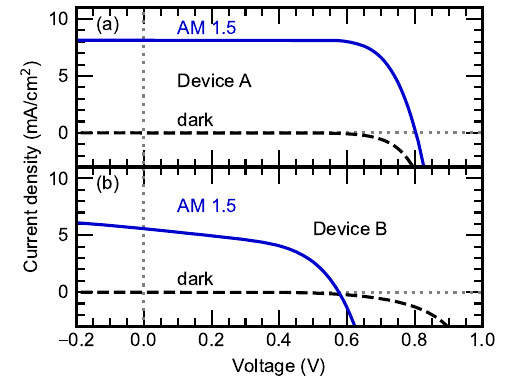}
    \caption{Current--voltage curves of (a) Device A and (b) Device B in the dark (dashed curves) and under 1 sun AM1.5G illumination (solid curves).}
    \label{fig:J_V}
\end{figure}

Figures~\ref{fig:J_V}(a) and (b) show the $J$--$V$ characteristics of Device A and Device B under 1 sun illumination, respectively. For Device A, we obtained a $J_\mathrm{SC}$ of 8.14 mA/cm$^2$, an open-circuit voltage ($V_\mathrm{OC}$) of 0.80 V, and a fill factor (FF) of 0.74, and for Device B, we obtained $J_\mathrm{SC}=5.58$ mA/cm$^2$, $V_\mathrm{OC}$ = 0.58 V, and FF = 0.51. The lower $J_\mathrm{SC}$ and $V_\mathrm{OC}$ values for Device B are consistent with the lower output current values observed in Fig.~\ref{fig:J_SC}(d), which are caused by suppressed thermal activation over the barrier at the heterointerface. The low contribution from thermal activation is partially compensated by intraband absorption of low-energy photons in the AM1.5 solar spectrum. From the superlinear increase in the output current shown in Fig.~\ref{fig:J_SC}, we expect that the effect of intraband absorption becomes more pronounced at higher excitation power densities, and thus Device B may outperform Device A under concentrated sunlight. Optimization of the heterointerface, the QDs, and the double-tunnel-junction structure may further increase the intraband absorption efficiency and lead to a $J_\mathrm{SC}$ that surpasses the $J_\mathrm{SC}$ of a conventional TPU-SCs with QDs even under 1 sun illumination.

\section{Single-color excitation using IR light}\label{app:Single_ex}
\begin{figure}[tpb]
    \includegraphics[width=\linewidth]{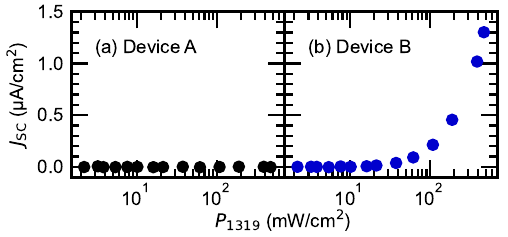}
    \caption{The $J_\mathrm{SC}$ of (a) Device A and (b) Device B exposed to IR light with a wavelength of 1319 nm.}
    \label{fig:ex_single}
\end{figure}

Figures~\ref{fig:ex_single}(a) and (b) show the dependence of $J_\mathrm{SC}$ on the IR excitation power density in the case of single-color excitation for Device A and Device B, respectively. Device A is insensitive to the 1319-nm light, even under the highest excitation power density of 480 mW/cm$^2$. On the other hand, we can confirm a clear increase in $J_\mathrm{SC}$ with $P_{1319}$ in the case of Device B. This difference is caused by the double-tunnel-junction structure. Although the detailed mechanism is still unclear, optical transitions from the VB in the degenerated $n^+$-layer to the CB of the degenerated p$^+$-layer including a bandgap shrinkage due to heavy doping may one possible explanation.

\section{Data of the short-circuit current for Device B}\label{app:J_SC_debB}
\begin{figure}[tpb]
    \includegraphics[width=\linewidth]{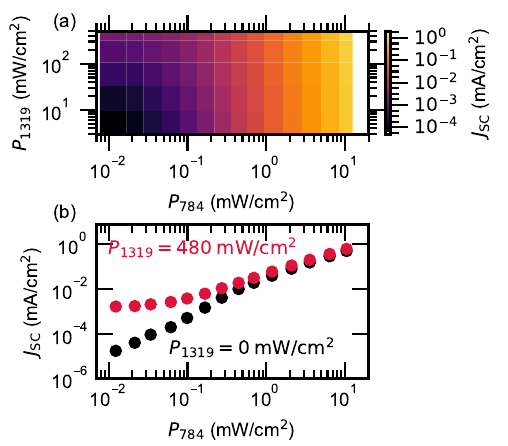}
    \caption{The behavior of the short-circuit current density of Device B without subtracting $J_\mathrm{SC\_1319}$: (a) The two-dimensional map of $J_\mathrm{SC}$ for all investigated two-color excitation conditions, and (b) the dependence of $J_\mathrm{SC}$ on $P_{784}$, where the black closed circles are the same data as in Fig.~\ref{fig:J_SC}(d) for $P_{1319}=$ = 0 mW/cm$^2$ and the red closed circles are the data for $P_{1319}=$ 480 mW/cm$^2$.}
    \label{fig:JSC_devB}
\end{figure}

Figure~\ref{fig:JSC_devB}(a) shows the two-dimensional map of the measured $J_\mathrm{SC}$ of Device B without subtracting $J_{\mathrm{SC}\_1319}$ as functions of $P_{784}$ and $P_{1319}$. Figure~\ref{fig:JSC_devB}(b) shows two data sets to illustrate the impact of the IR light on $J_\mathrm{SC}$. For $P_{1319}=$ 480 mW/cm$^2$, significant $J_\mathrm{SC}$ values are already observed under low 784-nm excitation power densities, which is caused by the interband transitions induced by the 1319 nm light. The effect of the interband transitions induced by the 1319-nm photons becomes less pronounced as $P_{784}$ is increased.

\section{Comparison of the experimental and fitted results}\label{app:sim_comp}
\begin{figure*}[tpb]
    \includegraphics[width=\linewidth]{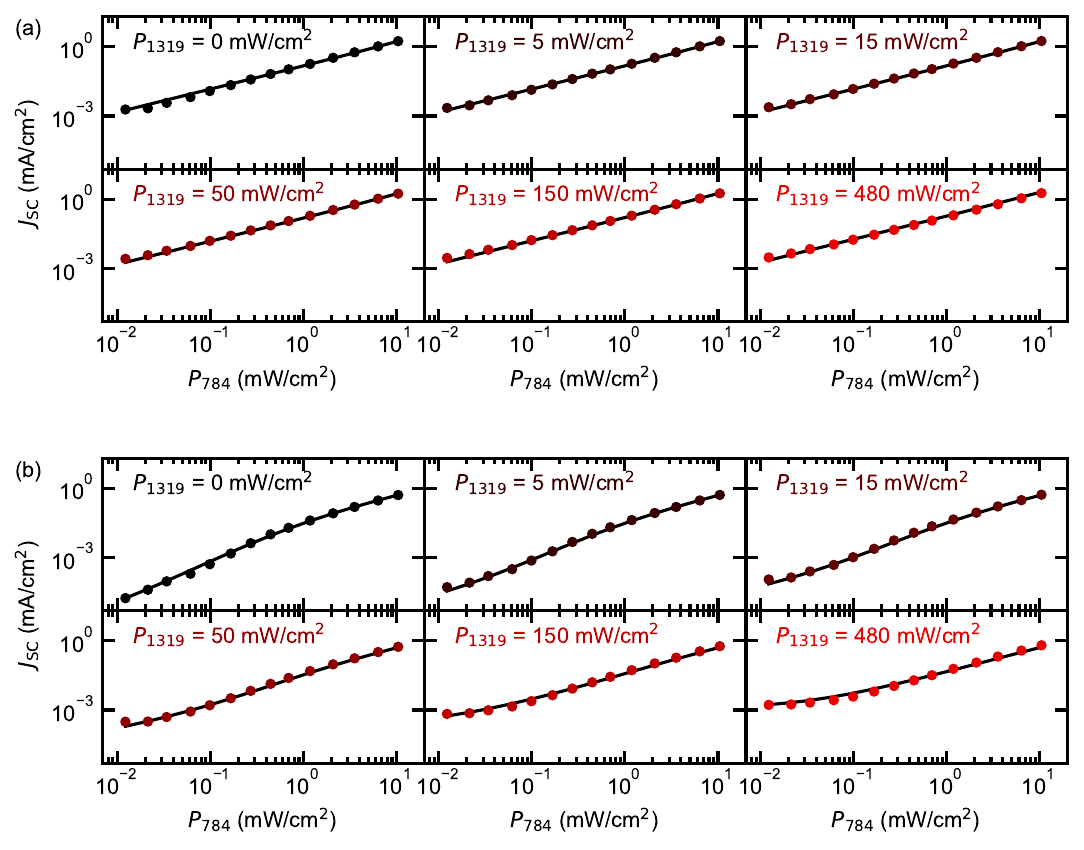}
    \caption{Comparison between the experimentally obtained $J_\mathrm{SC}$ values and the $J_\mathrm{SC}$ values obtained by global fitting of Eqs.~(\ref{eq:rate_1})--(\ref{eq:rate_4}) to all data sets. (a) and (b) show the results for Device A and Device B, respectively. The solid lines represent the results calculated using Eq.~(\ref{eq:calc_J}), and the closed circles are the experimental data.}
    \label{fig:sim_comp}
\end{figure*}

Figure~\ref{fig:sim_comp}(a) shows the experimentally obtained $J_\mathrm{SC}$ data sets and the $J_\mathrm{SC}$ calculated using Eq.~(\ref{eq:calc_J}) as a function of the 784-nm excitation power density for six different IR excitation power densities. The parameters used in the calculation are summarized in Table~\ref{tab:sim}. The calculated $J_\mathrm{SC}$  values accurately reproduce the experimental values, indicating that our model is able to explain the experimental observations for a rather wide range of conditions, even though simple rate equations were used.
\clearpage
\bibliography{references}

\end{document}